# Whitening the Sky: light pollution as a form of cultural genocide

**Duane W. Hamacher [1], Krystal De Napoli [2], and Bon Mott [3]**

[1] ASTRO-3D Centre of Excellence, School of Physics, University of Melbourne, Parkville, VIC, 3010, Australia.
[2] School of Physics & Astronomy, Monash University, Clayton VIC, 3080, Australia
[3] Faculty of Fine Arts & Music, University of Melbourne, Southbank, VIC, 3006, Australia.

\* Correspondence: duane.hamacher@unimelb.edu.au

**Abstract**: Light pollution is actively destroying our ability to see the stars. Many Indigenous traditions and knowledge systems around the world are based on the stars, and the peoples' ability to observe and interpret stellar positions and properties is of critical importance for daily life and cultural continuity. The erasure of the night sky acts to erase Indigenous connection to the stars, acting as a form of ongoing cultural and ecological genocide. Efforts to reduce, minimise, or eliminate light pollution are being achieved with varying degrees of success, but urban expansion, poor lighting design, and the increased use of blue-light emitting LEDs as a cost-effective solution is worsening problems related to human health, wildlife, and astronomical heritage for the benefit of capitalistic economic growth. We provide a brief overview of the issue, illustrating some of the important connections that the Aboriginal and Torres Strait Islander people of Australia maintain with the stars, as well as the impact growing light pollution has on this ancient knowledge. We propose a transdisciplinary approach to solving these issues, using a foundation based on Indigenous philosophies and decolonising methodologies.

## 1. Introduction

Cultures across space and time formed a close connection to the sky, whether it be through a philosophical, spiritual, and/or scientific perspective (Ruggles 2009; Ruggles 2015). Indigenous cultures in Australia, along with many cultures across the world, inextricably encompass all three perspectives (e.g. Clarke 2007). The stars are used to preserve and inform complex knowledge systems, which are used for things like navigation, food economics, forecasting weather, predicting seasonal change, informing social structure, and serving as a mnemonic for committing information to memory and passing it to successive generations over long periods of time (Hamacher 2012; Norris 2016; Kelly 2015).

In Australia, the cosmos serves as a foundation for numerous Indigenous knowledge systems and Origin stories (Tindale 2005). For Aboriginal and Torres Strait Islander people, the stars encode and communicate history, law, ethics, and moral values. Ngarinyin elder David Mowaljarlai said that "Everything is written twice – on the ground and in the sky" (Mowaljarlai and Malnic 1993), a description shared by Indigenous peoples around the world (e.g. Lee 2016). For example, the celestial emu is one of the most widespread Aboriginal asterisms across Australia (Fuller et al. 2014). It is not made up of the bright stars, but rather of the dark dust lanes in the Milky Way, between the Coalsack nebulae in Crux and the galactic centre in Scorpius and Sagittarius (Gullberg *et al*. 2020; Fig. 1). The visibility and position of the emu in the sky throughout the year informs Aboriginal people about the behaviour of the bird (referencing both sexes, depending on the time of year it is visible), the changing seasons, navigational pathways, and social practices (Fuller *et al*. 2013). This "dark constellation" - as well as the Magellanic Clouds, globular and open star clusters, nebulae, fainter stars, and aurorae - are being eradicated from view by increasing light pollution.

Traditionally, the concept of natural light pollution holds special significance to Aboriginal Australians. For example, the Gunnai people in east Gippsland, Victoria share oral traditions that describe the dynamic between the Moon man hunting the celestial emu. When the Moon rises in the sky, the glow of its light makes the emu "hide away" (Thorpe 2019), but it returns to visibility when the Moon sets. However, human-made light pollution is impacting the visibility of the emu and other fainter celestial objects altogether, making them impossible to see in urban areas, regional cities, and – increasingly – in remote communities.

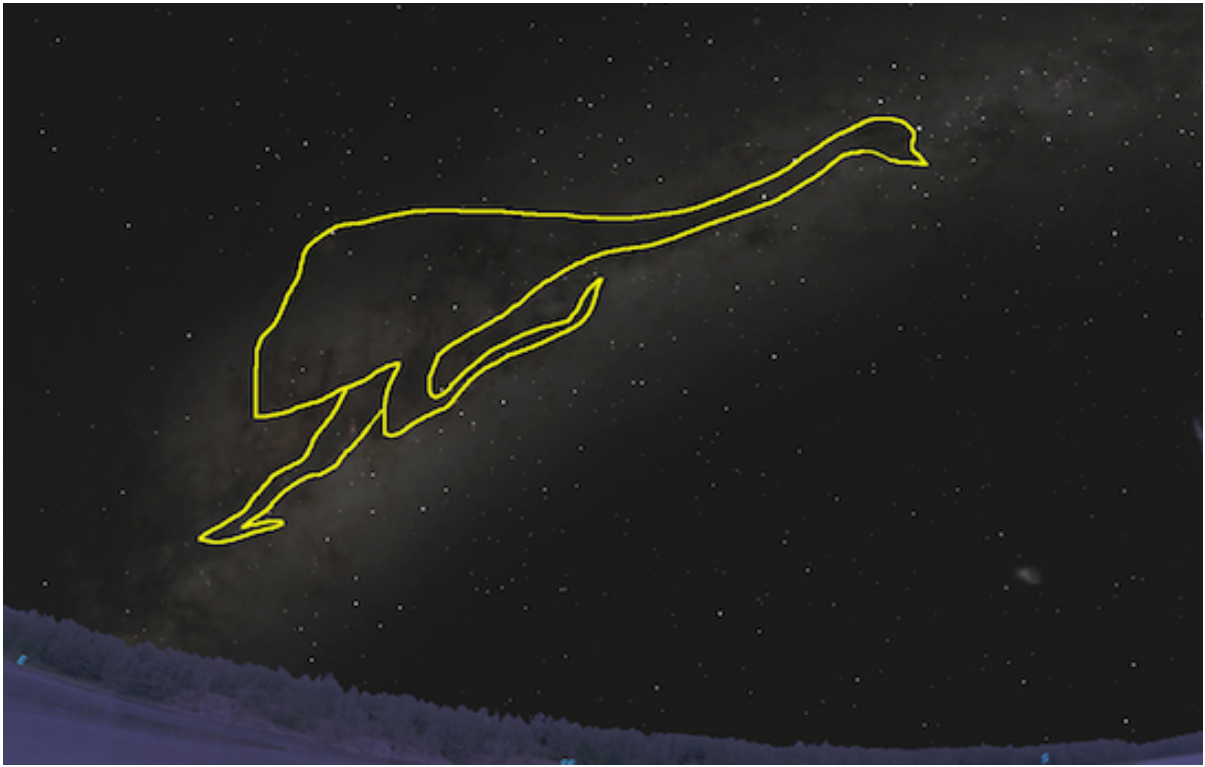

Fig. 1: The celestial emu, Gawarrgay, as described in Kamilaroi traditions of northern New South Wales, Australia. Image: Ghillar Michael Anderson and Robert Fuller, after Fuller *et al*. (2014a).

While our connection to the visible sky is fading under urban expansion and increased lighting (Falchi *et al*. 2016), steps are being taken to reconnect people to the inspiration of the cosmos and traditional knowledge of the stars. In the city of Melbourne, mosaics on popular walkways near Port Philip Bay feature the celestial emu, with information provided in plaques about its meaning in the local Boon Wurrung Aboriginal language (Fig. 2). While these educational means of communicating traditional knowledge are important, it is crucial to ensure that people can actually see the celestial emu in the sky itself.

2. **Preserving Dark Skies**

Aside from the general trend of urban expansion, one of the major issues causing excessive light pollution and associated negative impacts is the increased use of LED lighting in homes and businesses, streetlights, and car headlamps. In addition to the threat this places on Indigenous connections to the sky, increased lighting and the use of blue-rich LEDs, is having a negative impact on human health, wildlife, and our collective connection to the stars (e.g. Ticleanu and Littlefair 2015). For example, Ginan[1] (Epsilon Crucis), the fifth star of the Southern Cross – arguably the most well-known constellation in Australia – is barely visible (if at all) in the skies over major metropolitan areas. It is therefore of critical importance to ensure that skies be kept clear and dark to enable this connection to be maintained.

The development of Dark Sky preserves around the world is helping to moderate this issue. The Warrumbungle National Park, home to Australia's largest optical observatory at Siding Spring as well as Kamilaroi Aboriginal cultural sites linked to the Seven Sisters (Pleaides), is Australia's first registered Dark Sky Park (New South Wales Parks & Wildlife Services 2019). Despite this status, the area is under significant threat from the glow of distant cities and the large flares from coal-seam gas (fracking) in the Pilliga National Forest to the north (Fig. 3; Milman, 2014). Light pollution from fracking flares is an apt demonstration of ongoing theft and destruction of Aboriginal land leading to the erasure of the night sky for the financial gain of colonial interests.

---

[1] The name *Ginan* is a Wardaman Aboriginal name from the Northern Territory, meaning "dilly bag filled with songs of knowledge". It was approved as the official name of this star by the International Astronomical Union's Working Group on Star Names in 2017, of which the first author (DWH) is a member (Hamacher 2018).

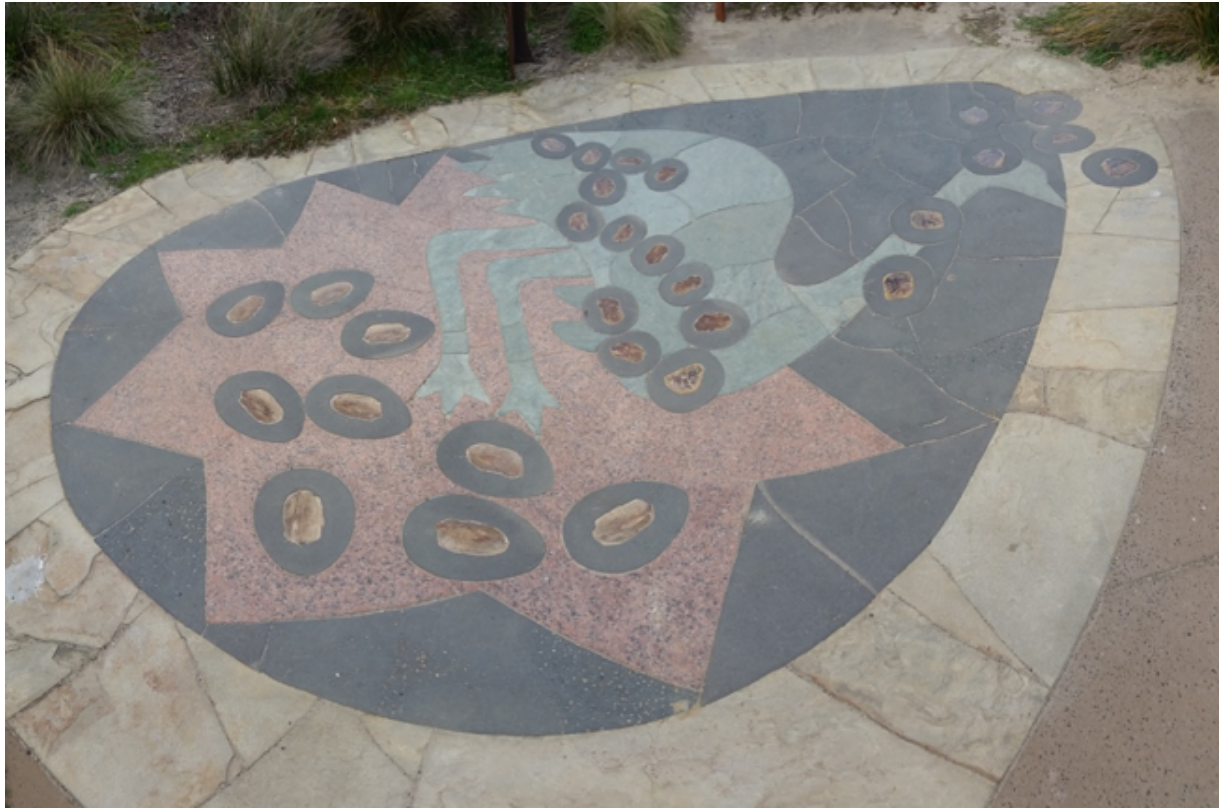

Fig. 2: The Boon Wurrung view of the emu in the sky, as a mural on a coastal pathway along Port Philip Bay near Brighton, Victoria. Image: D.W. Hamacher.

Dark sky parks set criteria to ensure dark skies, such as the modification of lighting so that it faces downwards, fitting lights with shields, replacing or disconnecting upward facing lights, placing lights in public areas on a timer to reduce the length of their use, and eliminating street lighting altogether in the Dark Sky Park areas (New South Wales Parks & Wildlife Services, 2019). While these measures are highly beneficial, an increasing number of metropolitan and regional councils, commercial organisations, and private citizens are switching to LED lights due to their low energy usage, low cost, and apparent "green" status. Because of their low cost, more of these lights are being used, causing a greater increase in light pollution (Dvorsky 2017).

Many of the most commonly used LED lights emit a substantial portion of their light in the blue end of the spectrum, which is damaging to the visibility of the sky and to the health of humans and wildlife (Royal Society Te Apārangi 2018). Organisations such as the Australian Dark Sky Alliance (www.australasiandarkskyalliance.org) and industrial lighting companies like WE-EF (www.weef.de) are working together to find solutions to this problem, which include the use of low-impact amber LEDs (which emit a majority of their light in the red/yellow end of the spectrum), covering street lights at the top to avoid direct exposure to the sky, angling street lights downward, and reducing overall light scatter (Elsahragty and Kim 2015).

The development of dark sky reserves is a great step towards raising awareness of the importance of preserving dark skies, particularly in rural areas. But light pollution is driven by economic growth and urban expansion (e.g. Chaiwat 2016), so substantial work is needed to improve and promote low-impact lighting in large population centers. An example of a community which has put a high priority on maintaining dark night skies is the city of Flagstaff, Arizona in the USA, which was designated a Dark Sky City in 2001 (City of Flagstaff, 2017). A major driving force behind this was the presence of numerous astronomical observatories in the area. The city's policies enable the general public to see dark, clear skies from an urban center - something not afforded to people in most other metropolitan areas. If substantial and sustainable progress is to be made in preserving dark skies, more cities will need to address the way they light their communities, as light pollution affects skies far outside the city limits.

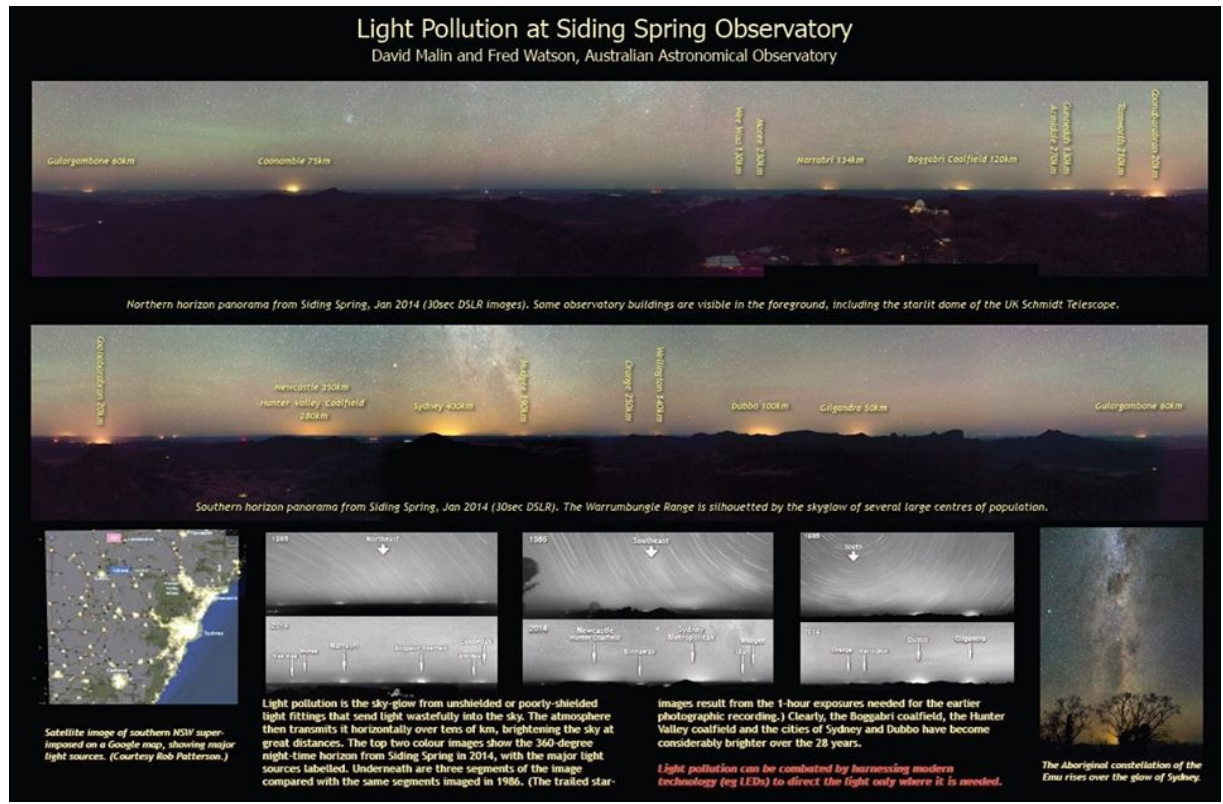

Fig. 3: Light pollution visible from Siding Spring Observatory over time, with increasing threats derived from regional cities and coal seam gas flares. Image: David Malin and Fred Watson, Australian Astronomical Observatory.

## 3    Finding Solutions Using a Trans-Disciplinary Approach

Addressing these issues poses a number of important questions: How can we maintain a connection to the stars when we cannot see them? How can we reduce or eliminate light pollution to preserve astronomical heritage and knowledge? How can we reduce the detrimental effects of excess light on the health of humans and wildlife? Accomplishing this requires solutions drawn from trans-disciplinary research, i.e. research that is conducted by investigators from different disciplines working jointly to create new conceptual, theoretical, methodological, and translational innovations that integrate and move beyond discipline-specific approaches to address a common problem (Aboelela *et al.* 2007). This research strategy creates a wholistic approach to scholarship intended to arrive at expanded research outcomes (Nicolescu 2008).

Transdisciplinary innovations related to Dark Sky Studies are driven by collaborations between astronomers, ecologists, engineers, industrial designers, heritage consultants, landscape architects, artists, and health professionals (see Ardavani et al. 2020), but rarely involve Indigenous communities. We argue that a decolonising approach must be adopted, using Indigenous theoretical and methodological foundations and philosophies (e.g. Tuhiwai-Smith 2013; Nakata 2002; Nakata 2010) in close-collaboration with Indigenous people, as opposed to focusing strictly on the "whitewashed" Western philosophies that dominate modern academic discourse (see Rutledge 2019). Doing so would reflect a more sustainable and humanistic approach.

With regard to the colonisation of Indigenous peoples and the degradation of Indigenous land (including water and sky), the expansion of light pollution, fuelled by industry and government, is arguably an ongoing continuation of cultural genocide – a concept often described as "slow violence". This can operate as a major threat multiplier by "fuelling long-term, proliferating conflicts in situations where the conditions for sustaining life become increasingly but gradually degraded" (Nixon 2011). Given that Indigenous cultures in places like Australia maintain close and essential connections with the stars, the very foundation of Indigenous cosmology, knowledge systems, social structure, and the library of oral traditions is being actively destroyed through encroaching light pollution. Additionally, LED lights emitting in the blue spectrum contribute to health issues by disturbing circadian rhythms (Akacem et al. 2016; Ayaki et al. 2016) and negatively impacting the health and behaviour of nocturnal wildlife (Altermatt and Ebert, 2016). This poses ethical concerns that must be addressed through government and industry practice.

Growing light pollution is damaging human and wildlife connections to the stars, emphasising a need for protecting and preserving dark skies. This is especially problematic in places where Indigenous people have maintained complex, deep-time knowledge systems in which the stars are encapsulated in their cosmologies and epistemologies. The whitewashing of the night sky through colonial policy and practice - without regard to Indigenous people, land, or culture - is an ongoing form of cultural genocide (Genocide Museum n.d.).

Preserving dark skies globally extends beyond benefits to economic growth, colonial interests, or scientific objectives (e.g. astrophysical observatories). It directly impacts the ongoing survival and prosperity of our human connection to the sky and the deep astronomical knowledge systems of the world's Indigenous cultures. Solutions to this must be transdisciplinary in nature and include Indigenous voices and philosophies that utilise a decolonising framework.

**Funding:** Duane Hamacher is funded by the Australian Research Council (Project DE140101600), and the Pierce Bequest and Laby Foundation in the School of Physics at the University of Melbourne.

**Acknowledgments:** Authors acknowledge the following for their input, influence, and support: Hilding Neilson, N'arweet Carolyn Briggs, Tomas Kanthak, Marnie Ogg, Jessica Heim, WE-EF, and the Australian Dark Sky Alliance.

**Conflicts of Interest:** Duane Hamacher is an Ambassador of the Australian Dark Sky Alliance and has been collaborating with WE-EF, although the company has not provided funding support for this research to date.

## About the Authors

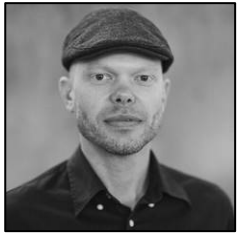

**Duane Hamacher**

Dr Duane Hamacher is Associate Professor of Cultural Astronomy in the ASTRO-3D Centre of Excellence in the School of Physics at the University of Melbourne, and an Ambassador of the Australian Dark Sky Alliance.

https://unimelb.academia.edu/DuaneHamacher

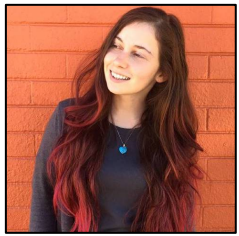

**Krystal de Napoli**

Krystal de Napoli is a proud Gomeroi Aboriginal woman pursuing a degree in astrophysics at Monash University in Melbourne, and a researcher with the CSIRO Data61 team.

https://www.linkedin.com/in/krystal-de-napoli

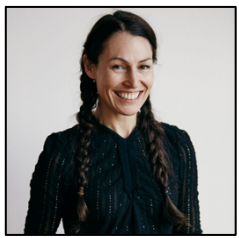

**Bon Mott**

Bon Mott is a sculptor, curator, and performance artist completing a PhD in Visual Art at the University of Melbourne, and Director of Second Space Studios (2SP) in Fitzroy.

http://www.bonmott.com